# The Chemical Reactions in Electrosprays of Water Do *Not* Always Correspond to Those at the Pristine Air-Water Interface

Adair Gallo Jr. [1,2,3], Andreia S. F. Farinha[1,2,3], Miguel Dinis[1,4], Abdul-Hamid Emwas[1,5], Robert J. Nielsen[6], William A. Goddard III[6], Himanshu Mishra[1,2,3,*]

[1]King Abdullah University of Science and Technology (KAUST),
[2]Water Desalination and Reuse Center (WDRC),
[3]Division of Biological and Environmental Sciences (BESE),
[4]KAUST Catalysis Center (KCC),
[5]Imaging and Characterization Core Laboratory,
Thuwal 23955-6900, Saudi Arabia;
[6]Materials and Process Simulation Center, California Institute of Technology,
Pasadena, CA 91125, USA
* Himanshu.Mishra@Kaust.edu.sa

## Abstract

The recent application of electrosprays to characterize the air-water interface, along with the reports on dramatically accelerated chemical reactions in aqueous electrosprays, have sparked a broad interest. Herein, we report on complementary laboratory and *in silico* experiments tracking the oligomerization of isoprene, an important biogenic gas, in electrosprays and isoprene-water emulsions to differentiate the contributions of interfacial effects from those of high voltages leading to charge-separation and concentration of reactants in the electrosprays. To this end, we employed electrospray ionization mass spectrometry, proton nuclear magnetic resonance, and quantum mechanical simulations. We found that the oligomerization of isoprene in aqueous electrosprays involved minimally hydrated and highly reactive hydronium ions. Those conditions, however, are non-existent at pristine air-water interfaces and oil-water emulsions under normal temperature and pressure. Thus, electrosprays should be complemented with surface-specific platforms and theoretical methods to reliably investigate chemistries at the pristine air-water interface.

# 1. Introduction

The air-water interface plays a critical role in numerous natural and applied contexts, such as atmospheric chemistries[1], precipitation[2], spray coatings[3], and materials synthesis[4, 5]. Indeed, it has been hypothesized that microdroplets generated during the splashing of waves in oceans could have been the chemical reactors leading to the origin of life[6-8]. Despite its ubiquity and importance, a variety of fundamental phenomena at the air-water interface remain incompletely understood, such as the specific adsorption of ions[9-11] and chemistries therein[8, 12-18]. The interfacial region, with a typical thickness $\delta_o \approx 0.5 \text{ nm}$, separates the gas-phase (vapor) from the condensed phase (water), two drastically different regions in terms of hydration - reactions spontaneous in one phase are forbidden in the other[19]. In fact, the chemical activities of species at the air-water interface can depart significantly from those in the bulk, as has been demonstrated by surface-specific techniques, including vibrational second harmonic generation and sum frequency generation[18, 20, 21], and polarization-modulated infrared absorption reflection spectroscopy[8, 11], and indirect approaches, including NMR[22] and confocal fluorescence microscopy[23]. Even though vibrational spectroscopy-based techniques report directly on thermodynamic properties of the air-water interface, they suffer from interpretational ambiguities and limitations due to low signal-to-noise ratios[24-30]. Thus, new techniques with higher sensitivity and unambiguous response are needed to help resolve the poorly understood features of the air-water interface while providing benchmarks to judge previous interpretations[31]. In this work, we assess the application of electrospray ionization mass spectrometry (ESIMS) to unravel the thermodynamic properties of pristine air-water interface (Henceforth, we will use the qualifier '**pristine**' to refer to the air-water interface that is not under the influence of any external sources/agents, such as electrical voltage or a drying gas).

In the recent years, ESIMS, which has been widely used to characterize ionic/molecular species in polar/apolar solvents[32], has been adapted to investigate the pristine air-water interface. In the standard configuration, ESIMS experiments entail the formation of electrosprays by the application of electrical potential and/or pneumatic pressure leading microscale droplets with excess electrical charge; those microdroplets pass through a glass/metallic capillary maintained at elevated temperature (~473 K) to evaporate the solvent and facilitate the mass spectrometric detection of analytes downstream[32-36]. In the experiments designed to investigate chemistries at the air-water interface, electrosprays containing one or more reactant(s) are intersected with gases or other electrosprays containing other reactant(s) followed by mass spectrometric detection. For instance, using this platform, thermodynamic properties of the pristine air-water interface have been explored under ambient conditions, including the relative concentrations of interfacial hydronium and hydroxide ions and their activities[37-41] leading to interpretations that have elicited scientific debate[9, 10, 15-17, 42, 43]. Further, by intersecting electrosprays of pH-adjusted water with gaseous organic acids[38], isoprene[39], and terpenes[38, 39, 44, 45], researchers observed instantaneous protonation (< 1 ms), and in some cases oligomerization of organics, which led them to conclude that as the bulk acidity of water approaches pH ≤ 3.6, the pristine air-water interface behaves as a superacid. While a clear understanding of the emergence of the putative superacidity at the air-water interface is unavailable, we note that in the condensed phase proton-catalyzed oligomerization of isoprene (or olefins in general) requires 60-80% concentrated $H_2SO_4$ solutions (pH < -0.5)[46]. Similar rate enhancements in aqueous electrosprays have also been observed for the syntheses of abiotic sugar phosphates[13, 47], the Pomeranz-Fritsch synthesis of isoquinoline[48], the reaction between o-phthalaldehyde and alanine[49], and the ozonation of oleic acid[50], among others[14,51]. Herein, we assess the relationships between the chemistries observed in aqueous electrosprays to those at pristine air-water interfaces; we also seek to decouple the factors that contribute to the mechanisms underlying reported dramatic rate enhancements by addressing the following questions:

(i) Do aqueous electrospray-based platforms report on thermodynamic properties of the **pristine** air-water interface?

(ii) Do accelerated reactions in aqueous electrosprays arise only from the significant enhancement of the hydrophobe-water (air-water) interfacial area? If yes, the mechanisms underlying the dramatic rate enhancements therein should be insightful in explaining the accelerated organic reactions in oil-water emulsions also referred to as 'on-water' catalysis[52-55].

(iii) Are the rate accelerations in aqueous electrosprays driven solely by the non-equilibrium conditions therein, especially the enhanced concentration of reactants in the micro-/nano-droplets due to the evaporation of water[56-59]?

(iv) Are gas-phase reactions implicated in the acceleration of chemical reactions in aqueous electrosprays[34, 42, 48, 60-62]?

To address those questions, we investigated the oligomerization of isoprene by proton nuclear magnetic resonance ($^1$H-NMR), a non-invasive technique, as a complementary platform to the ESIMS. Questions (i-ii) were addressed by comparing the effects of enhancing the water-hydrophobe interfacial area in both liquid-vapor and liquid-liquid systems; questions (iii-iv) were addressed by varying the capillary voltages, ionic strengths of the aqueous solutions electrosprayed and intersected with gas-phase isoprene, and $^1$H-NMR analysis of condensed vapor from the electrosprays. To highlight the role of hydration in electrosprays, we performed quantum mechanical calculations employing density functional theory (M0-6 flavor).

## 2. Materials and Methods

In our experiments, we used isoprene (99% purity from Sigma-Aldrich), Mili-Q deionized water (18 M$\Omega$-m resistivity), $D_2O$ (99.9% purity from Sigma Aldrich), ethanol (absolute from Merck Millipore), acetone (HPLC standard from VWR Chemicals), NaCl (>99% purity from Sigma Aldrich), HCl (36.9% concentration from Fisher Scientific), DCl (35% concentration 99% deuterium purity from Sigma Aldrich), and NaOH (>97% purity from Sigma Aldrich) to adjust the pH and ionic strengths.

**ESIMS:** All experiments were conducted in a commercial Thermo Scientific – LCQ Fleet electrospray ionization mass spectrometer in the positive ion mode, where a DC potential of 6-8 kV was applied to the needle, the tube lens voltage was 75 V, the sheath gas flow rate was 10 arb, the pressure was 1.2 torr at the convection gauge and $0.8 \times 10^{-5}$ torr at the ion-gauge, the flow rates of analytes were controlled by a calibrated syringe pump and ranged between 1-10 $\mu$L/min, the distances from the ion source and the inlet to the mass spectrometer were ~2 cm, and the distance between the electrospray and the tube ejecting isoprene was 1 cm.

**$^1$H-NMR:** All NMR spectra were acquired using a Bruker 700 AVANAC III spectrometer equipped with a Bruker CP TCI multinuclear CryoProbe (BrukerBioSpin, Rheinstetten, Germany); Bruker Topspin 2.1 software was used to collect and analyze the data. We transferred 100 $\mu$l of the (**A1**) samples into 5 mm NMR tubes, followed by 600 $\mu$l of deuterated chloroform (CDCl$_3$). The $^1$H-NMR spectra were recorded at 298 K by collecting 32 scans with a recycle delay of 5 s, using a standard 1D 90$^o$ pulse sequence and standard (zg) program from the Bruker pulse library. The chemical shifts were adjusted using tetramethylsilane (TMS) as an internal chemical shift reference. The (**A**) samples and a sample of as-purchased isoprene (**B**) were prepared by transferring 100 $\mu$l of each to 5 mm NMR tubes, and then adding 550 $\mu$l of deuterated water D$_2$O to the NMR tubes. The $^1$H-NMR spectra were recorded by collecting 512 scans with a recycle delay time of 5 s, using an excitation sculpting pulse sequence (zgesgp) program from the Bruker pulse library. The chemical shifts were adjusted using 3-Trimethylsilylpropane sulfonic acid (DSS) as an internal chemical shift reference. The free induction decay (FID) data were collected at a spectral width of 16 ppm into 64k data points. The FID signals were amplified by an exponential line-broadening factor of 1 Hz before Fourier transformation.

**Computational methods:** Here we used the M06 family of DFT with the geometries minimized using the 6-311G** basis set for H, C, and O atoms[61]. Then we carried out single-point electronic energy calculations, $E_{\text{elec}}$, including the diffuse 6-311G**++ basis-set for all atoms[62]. The Hessians at these geometries were calculated to determine that the minima and transition states led to 0 and 1 imaginary frequency, respectively. The transition state structures were obtained by following the steep ascent or descent along the vibration mode with one imaginary frequency until the saddle point was reached. The vibrational frequencies from the Hessians were also used to provide the zero-point energies and vibrational contributions to the enthalpies and entropies. The free energies of isoprene at 1 atm were calculated using statistical mechanics for ideal gases. Under our experimental conditions distances larger than the sizes of our clusters separated the ions, so we excluded counter-ions from the simulations.

# 3. Results

We investigated chemical reactions between pH-adjusted water and isoprene ($C_5H_8$, 2-methyl-l,3-butadiene, MW = 68 amu, and solubility in water, $S$ = 0.7 g/L at normal temperature and pressure (NTP): 293 K and 1 atm). We chose to examine reactions of isoprene because (i) we wanted to reproduce previous experimental results to ensure a clear comparison, (ii) isoprene is an important biogenic gas whose fate in the atmosphere is not completely understood[1, 39, 63, 64], and (iii) we could investigate chemistries in electrosprays and emulsions by taking advantage of the low boiling point of isoprene ($T_b$ = 307 K) and the high vapor pressure at NTP ($p$ = 61 kPa)[65].

As delineated in Figure 1 and summarized in Table 1, we report on the following sets of ESIMS (detection limit = ~ 1 nM) and [1]H-NMR (detection limit = ~ 10 µM) experiments:

**(A) Liquid-liquid collisions**: At NTP, we combined liquid isoprene with pH-adjusted $H_2O$ or $D_2O$, $1 \leq pH \leq 13$, in a volumetric ratio 1:6:3 (isoprene:water:air), agitated the emulsions at 1200 rpm in a vortexer for 6, 60, or 360 minutes, and analyzed the organic phases after phase-separation by ESIMS and [1]H-NMR. Since the air in the reaction vessels was saturated with isoprene, those experiments also ensured the presence of the products of reactions between the gas-phase isoprene and pH-adjusted water in the organic phase.

**(A1) Condensed vapors from electrosprays of organic phase from (A)**: After the liquid-liquid collision reactions (**A**) were over and the organic phases were electrosprayed in ESIMS for characterization, we condensed the sprays and analyzed them by [1]H-NMR. The [1]H-NMR-based investigation of the reaction products from experiments (**A**) before and after electrospraying was carried out to pinpoint the effects, if any, of electrospraying on the formation of the products.

**(B) Pure components**: we analyzed as-purchased isoprene, acetone, and ethanol by ESIMS and [1]H-NMR.

**(C) Gas-liquid collisions**: we created electrosprays of aqueous solutions with varying ionic strengths and pH, and intersected them with a stream of gas-phase isoprene (0.48 g/min carried by $N_2$ gas flowing at 600 mL/min, i.e. isoprene gas concentration was 800 mg/L) followed by mass spectrometric detection (Methods).

Hereafter, throughout the paper, we will refer to our experiments on the **liquid-liquid collisions as (A), condensed vapors from the electrosprays (A1), pure isoprene as (B), and gas-liquid collisions in the ESIMS as (C)** (Figure 1, Table 1).

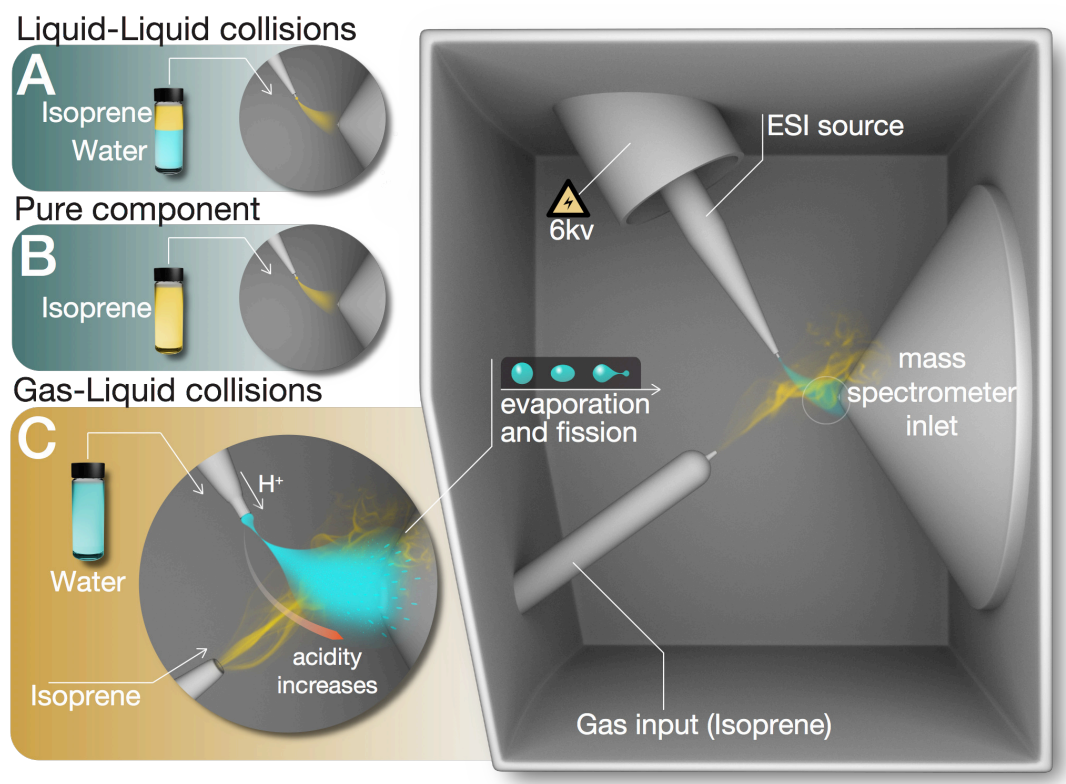

**Figure 1 – Summary of the experiments (A), (B), and (C) reported in this work along with three possible mechanisms for the oligomerization of isoprene during gas-liquid collisions.** (**A**) Liquid-liquid collisions: mixtures of isoprene, pH-adjusted water, and air in the volumetric ratio 1:6:3 was agitated at 1200 rpm (for 6, 60, 360 minutes) followed by ESIMS analysis of the organic phase. (**B**) As-purchased liquid isoprene was injected directly in the ESIMS. (**C**) Gas-liquid collisions: electrosprays of water (pH range 1-13) were collided with a stream of air carrying isoprene gas, followed by mass spectrometric detection (Methods).

## Table 1 – Experimental summary

| | (**A**) Liquid-liquid collisions | (**B**) Pure components | (**C**) Gas-liquid collisions |
|---|---|---|---|
| | Water(L)-Isoprene(L) | Isoprene(L), Acetone(L), Ethanol(L) | Isoprene(G)-Water(L) |
| **ESIMS** | Organics injected | Components injected | Water injected |
| pH | 1-13 | - | 1-13 |
| pNaCl | - | - | 1-9 |
| Shaking time | 6, 60, 360 min. | - | - |
| Voltage | 6 kV | 6 kV | 6-8 kV |
| Capillary temperature | 150 °C | 30-330 °C | 150 °C |
| **$^1$H-NMR** | Organics from (**A**) and condensed vapors (**A1**) | Isoprene(L) | - |
| pH | 1.5 | - | - |
| Shaking time | 6, 60, 360 min. | - | - |
| Aqueous phase | D$_2$O, H$_2$O | - | - |

Intriguingly, the ESIMS spectra from the above-mentioned experiments (A) at pH = 1, (B), and (C) at pH = 1 were nearly identical after normalizing with the maximum intensity (Figure 2A-C and Section $S_a$). The positions of the main peaks in the mass spectra fitted the general formula, $[(Isop)_n.H]^+$, which corresponded to covalently bonded oligomers of isoprene with one excess proton. In Section $S_b$ and Figure S1 we present the evidence proving that the peaks did not correspond to physisorbed clusters. In experiments (A), the mass spectra remained the same as the duration of mixing varied from 6 min to 6 hrs (Figure S2). We also observed numerous secondary peaks between the primary $[(Isop)_n.H]^+$ peaks that corresponded to the partial fragmentation of the four carbon bonds in the isoprene molecule ($C_5H_8$) and subsequent oligomerization of the fragments.

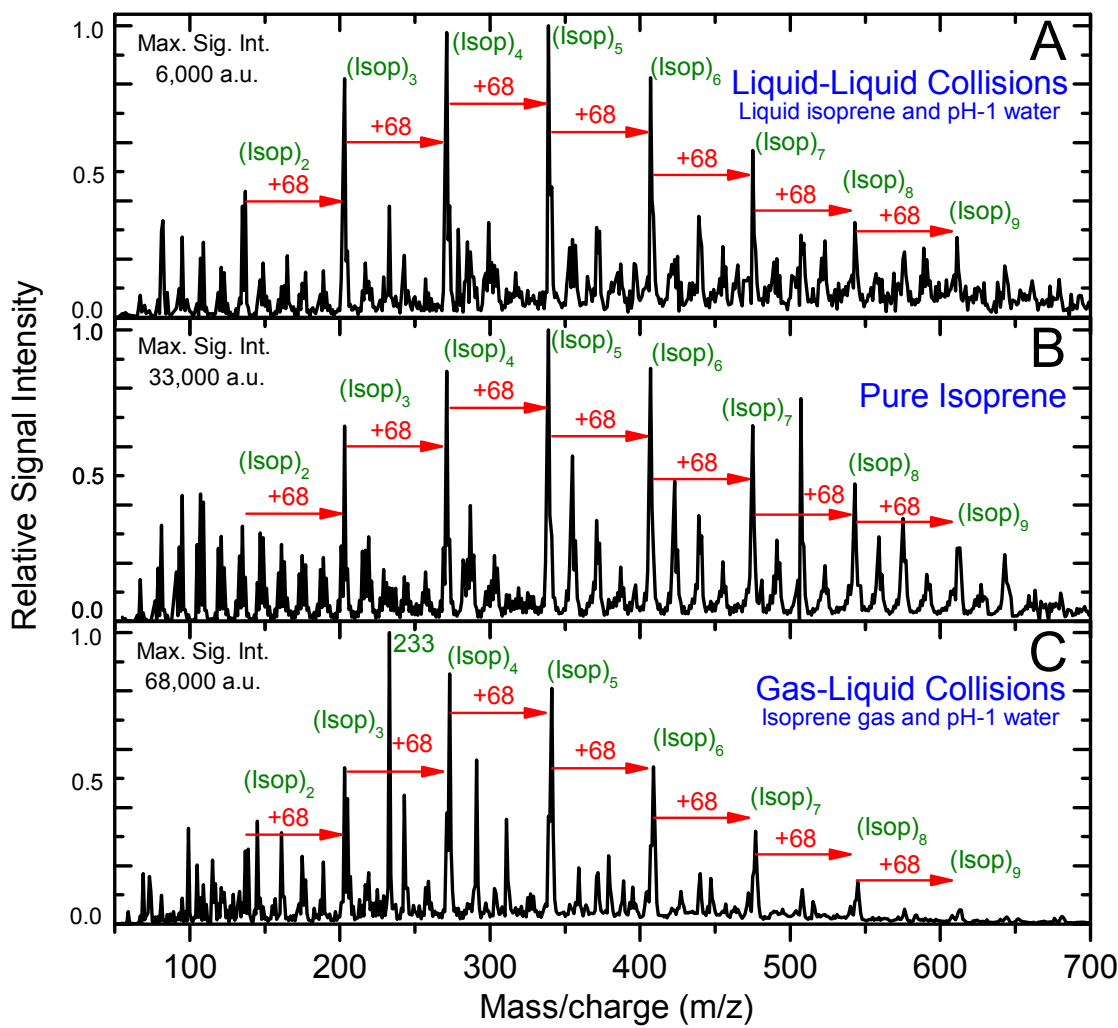

**Figure 2 – ESIMS spectra for sets of experiments A, B, and C:** The dominant peaks correspond to protonated oligomers of isoprene, $[(Isop)_n.H]^+$, and the secondary peaks correspond to fragments of the isoprene molecules attached to the primary peaks. (**A**) ESIMS spectra of the organic phase from the emulsion of liquid isoprene in water at pH = 1 and air (1:6:3 v/v/v) that was agitated at 1200 rpm for 360 minutes. (**B**) ESIMS spectra of as-purchased liquid isoprene. (**C**) ESIMS spectra of products of gas-liquid collisions between water (pH = 1) and gas-phase isoprene (Methods).

Next, we investigated the role of water pH on the oligomerization of isoprene in experiments (A) and (C). When the products were characterized by ESIMS, we noticed that the oligomers $[(Isop)_n.H]^+$ appeared only when the aqueous phase had pH ≤ 3.6 (Figure 3). Those observations have been reported previously[16, 38, 39, 44, 45] and ascribed to the superacidity of the air-water interface at pH ≤ 3.6. However, we also found that the ESIMS spectra from both experiments, (A) and (C), yielded oligomers $[(Isop)_n.H]^+$ for the acidic, basic, and pH-neutral salty solutions (Figure 3; compare Figure 2A, C with Figure S3 panels C1, C2, and C3).

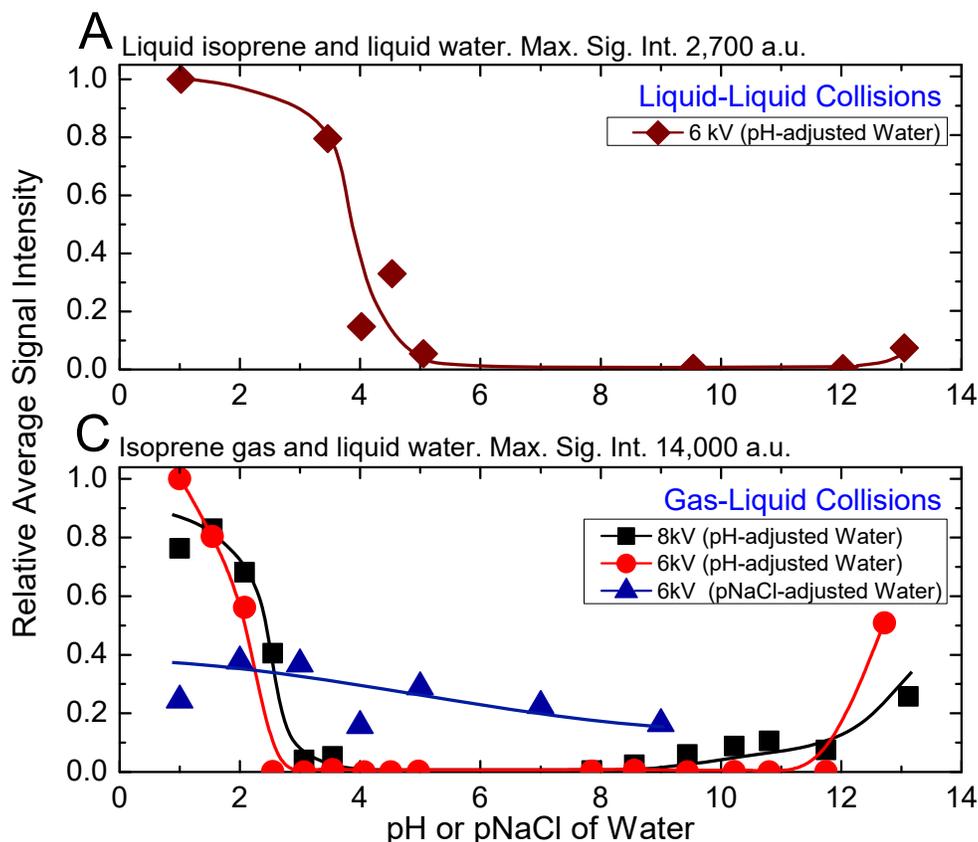

**Figure 3** – Influence of the ionic strength and ESI voltage on the ESIMS spectra of experimental sets (**A**) and (**C**). On the y-axis, we plot the average mass spectral intensity of all the oligomeric peaks $[(Isop)_n.H]^+$, given by $\sum_n I_n/n$, normalized by the highest datum in each plot. (**A**) Liquid-liquid collisions: the ESIMS spectra demonstrated protonation and oligomerization of isoprene after emulsions of isoprene in water with pH ≤ 3.6 and pH > 12 and air in a 1:6:3 ratio (v/v/v) were agitated at 1200 rpm for 360 minutes. (**C**) Gas-liquid collisions: the ESIMS spectra demonstrated protonation and oligomerization of isoprene gas after collision with electrosprays of water with pH ≤ 3.6 and pH > 12, and pH-neutral salty solutions. Curves are added to the plots to aid visualization.

In experiments (A), after the emulsions comprising liquid isoprene, liquid water at pH = 1.5, and air (containing saturated gaseous isoprene) were vigorously mixed (for 6 min, 60 min, and 360 min) we compared the organic layers after phase separation by ¹H-NMR. We also recorded the ¹H-NMR spectra of pure, as-purchased isoprene (B). To our surprise, the ¹H-NMR spectra from all of the set (A) samples were identical to those of set (B), indicating that the effect of the duration of shaking (6 – 360 min), the pH (1-13), the isotope (H₂O versus D₂O) and the presence of gaseous isoprene colliding with pH-adjusted water **did not lead to any oligomers** within the detection limit of 10 μM (Figure 4A and B). To investigate further, we condensed the vapors from the ESIMS exhaust (A1) after injecting the set (A) samples (1 ≤ pH ≤ 13), and obtained their ¹H-NMR spectra. All of (A1) samples showed spectra similar to each other (Figure 4A1). The ¹H-NMR spectra of (A) and (B) showed no sign of oligomers in the products: they contained a singlet at 1.87 ppm due to the resonance of the 3 protons in CH₃; three dublets at 5.02 ppm due to the resonance of the two protons H5b and H5b (coupling constant, $J$ = 13.2 Hz), 5.09 ppm due to the resonance of H1a with a cis-coupling constant, $J$ = 10.8 Hz; a dublet at 5.20 ppm due to the resonance of H1b with a trans-coupling constant of $J$ = 17.5 Hz; and two dublets at 6.47 ppm due to the resonance of the protons H2a and H2b with the corresponding trans- and cis-coupling constants, $J$ = 17.5 Hz and $J$ = 10.8 Hz. In contrast, the ¹H-NMR spectra of the condensed vapors from the electrosprays (A1) of the organic phase after the liquid-liquid collisions (A) demonstrated a dramatic increase in the complexity of the spectrum[66], indicating that the protonation and oligomerization of isoprene took place exclusively in the electrosprays.

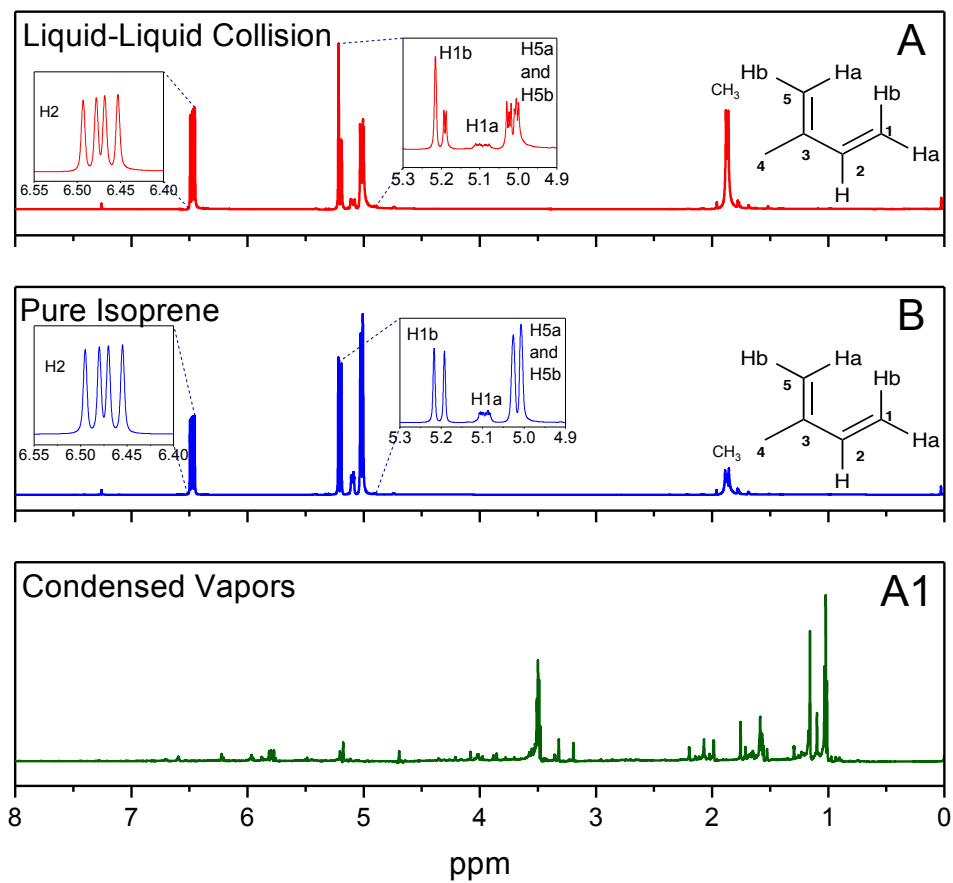

**Figure 4** − (**A**) $^1$H-NMR spectra of the organic phase after shaking liquid isoprene with pH 1.5 water and air in a volumetric ratio of 1:6:3 for 60 minutes. (**B**) $^1$H-NMR spectra of as-purchased isoprene. (**A1**) $^1$H-NMR spectra of the condensed exhaust from the electrosprays of the organic phase after the liquid-liquid collision (A) experiments. The nearly identical spectra for experimental sets (A) and (B) demonstrate that there was no detectable oligomerization of isoprene during the vigorous shaking of emulsions comprised of liquid isoprene with pH 1.5 water and air in a volumetric ratio of 1:6:3 (NMR resolution ~ 10 μM).

## 4. Discussion

Our investigation of experiments (A) with $^1$H-NMR revealed that a significant enhancement in the hydrophobe-water surface area was not sufficient for observable rate accelerations in emulsions of isoprene (gas and liquid) with pH-adjusted water at NTP conditions. On the other hand, analysis of experiments (A), (B), and (C) by ESIMS and experiments (A1) with $^1$H-NMR unambiguously demonstrated that *the chemical reactions took place* **exclusively** *in aqueous electrosprays* – the acidity, basicity, and saltiness of water all promoted the reactions. Further, as the capillary voltage was increased from 6 kV to 8 kV, the inflection points in experiments (A) and (C) shifted such that the oligomers $[(Isop)_n.H]^+$ were detected at lower ionic strengths (Figure 3C). Collectively, **these findings contradict previous claims of 'superacidity' of pristine air-water interfaces at pH ≤ 3.6.**

Next, we sought to identify the mechanisms underlying the protonation and oligomerization of isoprene in electrosprays (experiments C). As discussed above, a variety of parameters could influence reactions therein, including electrical voltage, electrochemical reactions, concentration of reactants in rapidly evaporating drops, and gas-phase reactions[32-34, 42, 48, 61, 62, 69-73]. Interestingly, by monitoring the changes in the surface tension of pendant water drops exposed to isoprene gas, we found that **gas-phase isoprene molecules could adsorb at the air-water interface** under NTP conditions (Figure S4). While the adsorption of non-polar molecules at the air-water interface might appear unexpected, similar phenomenon at the macroscale, entailing the adsorption of hydrophobic particles onto water drops of size $10^{-3}$ m in air forming 'liquid marbles' is well known[74]. Thus, gas-phase isoprene molecules (partial pressure in our chamber: 0.28 atm) may adsorb onto the positively charged aqueous electrosprays comprising excess protons[60]. From this stance, three potential mechanisms for the oligomerization of isoprene emerge, which we discuss and evaluate based on our experimental results and quantum mechanical predictions (Figure 1): **Mechanism M$_1$** – the adsorption of isoprene molecules onto the electrosprays increases their concentration at the interface, leading to reactions under the influence of high electric fields, similarly to the oligomerization of pure liquid isoprene on injection into ESIMS (Figure 2B); **Mechanism M$_2$** – continuous evaporation of positively charged electrosprays renders them increasingly acidic, akin to 50% $H_2SO_4$ solutions[46], which drives the liquid-phase oligomerization of the adsorbed isoprene molecules (Section S$_c$ and Figure S5); **Mechanism M$_3$** – molecular clusters of water molecules with an excess proton are ejected during Coulomb explosions, which protonate and oligomerize isoprene molecules in the gas-phase (Theoretical simulations section, Section S$_c$, and Figure S5). Mechanism M$_3$ is similar to that of proton transfer reaction mass spectrometry (PTRMS), which has been exploited to detect trace gases in the atmosphere[75]. In all those mechanisms, the initial ionic strength of water (acidic, basic, or salty) and electrical voltage were crucial for the formation of a stable stream of charged microdroplets - the higher the ionic strength of solutions, the lower the requirement for the electrical voltage[33, 34, 76-78]. In fact, due to the electrochemical reactions at the electrospray needle under the influence of high electric fields, the electrosprayed droplets from a positively charged capillary should contain more positive ions than in the bulk[56, 61, 77, 79, 80] (Section S$_c$, Figure S5). Interestingly, for pH-adjusted water electrosprayed at 6 kV, we detected oligomers (Figure 3C) when pH ≤ 3.6 or pH > 12, whereas for the NaCl solutions, we observed oligomers at concentrations as low as $10^{-9}$ M (pNa = 9). Yet, the higher intensities of the $[(Isop)_n.H]^+$ at pH ≤ 3.6 in comparison to the salty solutions (Figure 3C) indicate that the proposed **mechanism M$_1$ is unlikely to play a crucial role** in the case of gaseous isoprene interacting with electrosprays of water.

Following our logical exclusion of mechanism M$_1$, we are left with mechanisms M$_2$ and M$_3$, i.e. did the reactions take place on the surface of electrosprayed water droplets or in the gas-phase? Whether or not the electrospray spectra represent the solvent- or gas-phase chemistries/characteristics is a much-debated matter and case-specific[34, 36, 70, 81, 82]. Obviously, the answer would have a bearing on the questions (ii-iv) outlined above, because the kinetics and thermodynamics of reactions in bulk and gas-phase differ dramatically[19]. Recently, Silveira and co-workers employed cryogenic ion mobility mass spectrometry to demonstrate the effects of rapid dehydration on the structures of undecapeptide substance during the final stages of electrospray ionization[82]. Analogously, to gain insights into the role of hydration on protonation and oligomerization of isoprene in a model ESI process, we carried out quantum mechanics calculations (Computational Methods, Section $S_d$ and Figures S6-S8). To simulate the chemical reactions of isoprene in the electrosprays along the mechanism $M_3$, entailing gas-phase reactions, we investigated the interactions of a minimally hydrated cluster, $(H_2O)_3.H^+$, with an isoprene molecule. We found that they readily formed an adduct with the release of $\Delta H^0 = -14.2$ kcal-mol$^{-1}$ and $\Delta S^0 = -43.7$ cal-K$^{-1}$-mol$^{-1}$ from the loss of the translational and rotational degrees of freedom (Figure 5). The subsequent proton transfer was impeded by an easily surmountable barrier of $\Delta G^{\ddagger} = 5.8$ kcal-mol$^{-1}$ barrier, consistent with our ESIMS experiments. Further oligomerization with an additional isoprene molecule was impeded only by a barrier of $\Delta G^{\ddagger} = 2.1$ kcal mol$^{-1}$ (Figure 5). However, while working with a cluster of 36 water molecules and an excess proton, $(H_2O)_{35}\cdot H_3O^+$, to represent the pristine air-water interface, we found the kinetic barriers to protonation of isoprene to be $\Delta G^{\ddagger} = 25.5$ kcal-mol$^{-1}$, and the barrier to oligomerization, $\Delta G^{\ddagger} = 40.2$ kcal-mol$^{-1}$ (Figure 6). Those kinetic barriers could not be surmounted within 1 ms under NTP conditions, as evidenced by the $^1$H-NMR results (Figure 4 A, B). While our cluster of 36 water molecules is a crude approximation of the air-water interface, it qualitatively predicts that the oligomerization reactions would be severely impeded in larger clusters and drops of water as observed in our experiments with emulsions (observed by $^1$H-NMR). Thus, our simple quantum mechanical models predicted that the protonation and oligomerization reactions in electrosprays involved minimally hydrated hydronium ions (Mechanism M3), such as formed during Coulomb explosions, which are unavailable at the pristine surface of mildly acidic water under NTP conditions.

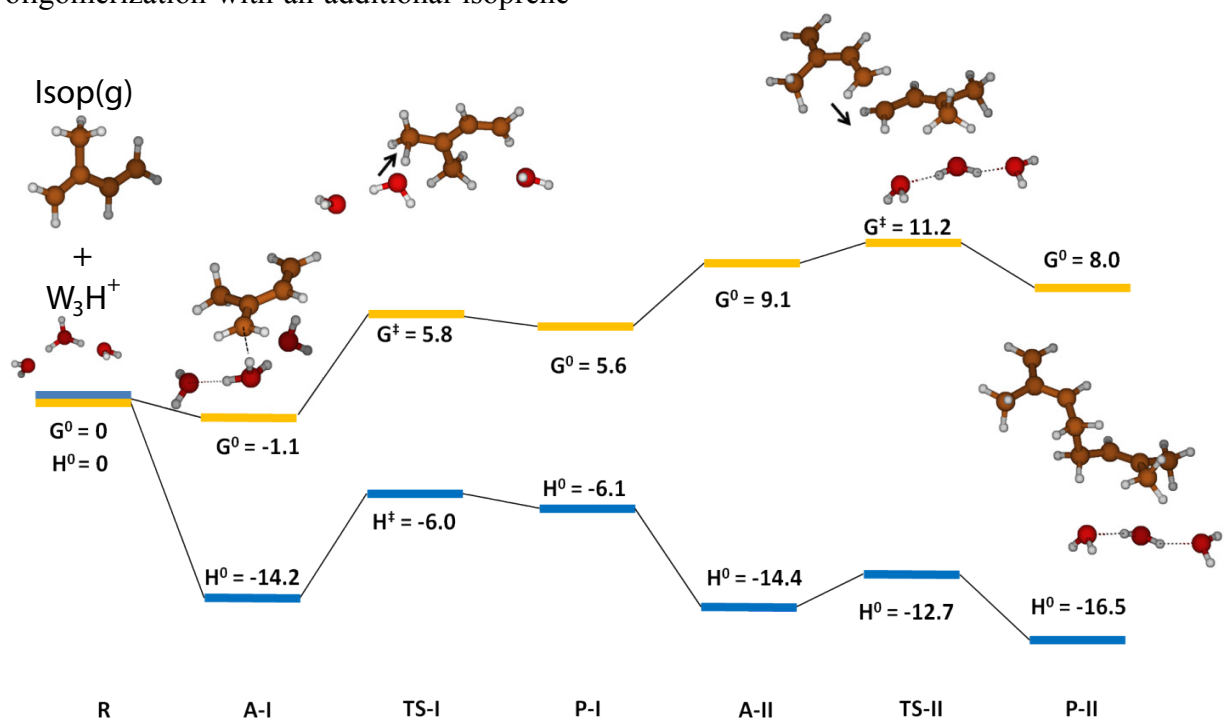

**Figure 5** − Quantum mechanics-based free energy and enthalpy landscapes for protonation and oligomerization of isoprene while interacting with a cluster comprising three water molecules and one excess proton, $(H_2O)_3.H^+$. Due to its incomplete hydration (compared to bulk), the proton exhibited extreme acidity. The free energy barrier for the proton transfer from $(H_2O)_3.H^+$ to isoprene(g) was $\Delta G^{\ddagger} = 6.9$ kcal mol$^{-1}$ and the barrier to subsequent oligomerization with another free isoprene(g) was $\Delta G^{\ddagger} = 2.1$ kcal mol$^{-1}$, which are easily surmountable under ambient NTP conditions within the timescale of our experiments (~1 ms). These model predictions support Mechanism M$_3$.

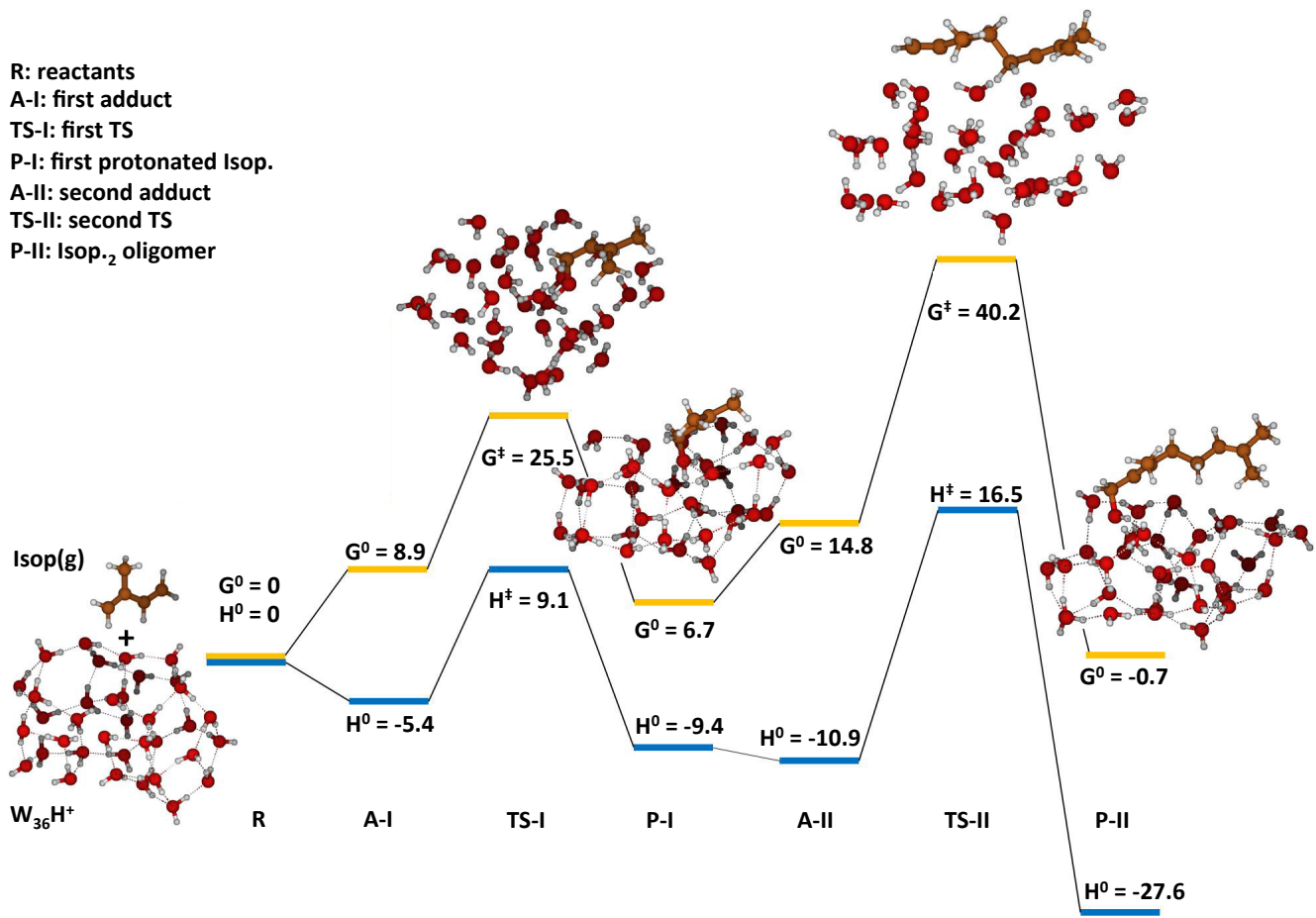

**Figure 6** – Quantum mechanics-based free energy and enthalpy landscapes for protonation and oligomerization of isoprene on a cluster consisting of 36 water molecules and an excess proton, $(H_2O)_{36}.H^+$, representative of very small water droplets. The kinetic barriers preventing proton transfer to isoprene and its subsequent oligomerization were $\Delta G^{\ddagger} = 25.5$ kcal mol$^{-1}$ and $\Delta G^{\ddagger} = 40.2$ kcal mol$^{-1}$, respectively which were insurmountable under ambient NTP conditions within the timescale of our experiments (~1 ms). The predictions of this model suggest that the reactions of isoprene in electrosprays cannot involve liquid-phase drops. These predictions also support the proposed mechanisms M$_3$ by ruling out the possible reactions through M$_2$.

# 5. Conclusions and Outlook

Based on our experimental investigation of oil-water and air-water interfaces of isoprene with pH-adjusted water, analyzed by ESIMS and [1]H-NMR along with quantum mechanical predictions, we address the questions outlined in the introduction as follows:

(i) Aqueous electrosprays need not always report on the thermodynamic properties of pristine air-water interfaces. Specially, for the case of analytes that may react under acidic conditions.

(ii) The observed chemical reactions of isoprene in aqueous electrosprays were not driven by the enhancement in the hydrophobe-water interfacial area, as evidenced by the lack thereof in vigorously mixed emulsions of isoprene and pH-adjusted water. Thus, the mechanisms underlying the 'on-water' catalysis[52-55] must be different from those leading to rate accelerations in the aqueous electrosprays[83]. Electrosprays of water must facilitate additional chemical pathways, such as reactions with partially hydrated (gas-phase) hydroniums, which are not accessible in vigorously mixed oil-water emulsions or pristine aqueous interfaces.

(iii) Reactions of isoprene in aqueous electrosprays were driven by non-equilibrium conditions therein - most importantly, due to the rapid evaporation of water leading to highly concentrated droplets and then to gas-phase hydroniums (**Mechanism M₃**). Other researchers have also found that the enhancement in the concentrations of reactants led to dramatic rate accelerations in electrosprays[48, 51, 69, 72, 84, 85]. Thus, the conditions of the aqueous solutions injected at ESIMS (pH, salinity) and the conditions of the set-up (voltage, temperature, auxiliary and sheath gas flows rates, etc.) significantly affect the formation of electrosprays and subsequent reactions.

(iv) Gas-phase reactions could play a significant role in the electrosprays – in our experiments, reactions between partially hydrated protons and isoprene led to its protonation and oligomerization, as recently suggested by Yan & co-workers[51].

Our experimental and theoretical results demonstrate that chemistries in aqueous electrosprays do not necessarily correspond to those at the pristine air-water interface and oil-water emulsions at NTP. These findings also contradict the previous claims of the superacidity of the pristine air-water interface as the bulk acidity approaches pH $\leq 3.6$[38]; the proposal for the mildly acidic environmental surfaces to act as the primary sink for the atmospheric isoprene/terpenes should also be reevaluated[39, 44, 45]. While the potential of aqueous electrosprays to produce high-value products appears promising, those reactions are unlikely to be realized at pristine aqueous interfaces because of the seminal role of the non-equilibrium effects, such as the formation of water clusters with minimally hydrated hydronium ions. We do note that air-water interfaces could, perhaps, be investigated semi-quantitatively through electrosprays, if the reactants do not participate in gas-phase, acid catalyzed, or redox reactions therein[49, 86-88] and/or the gas-phase reactants do not dissolve in the droplets to re-emerge as interfacial species; a careful case-by-case assessment is warranted. We conclude by stressing on the importance of combining complementary experimental techniques and molecular simulations in the quest to unravel phenomena occurring at the pristine air-water interface.

**Acknowledgements:** The research reported in this publication was supported by funding from King Abdullah University of Science and Technology (#OSR-2016-CRG5-2992). The authors thank Mr. Ivan Gromicho, Scientific Illustrator at KAUST, for preparing Figure 1. The authors also thank Professor Richard Saykally and Professor Evan Williams (University of California Berkeley) for fruitful discussions.

# Supporting Information

**Section S$_a$:**

**Comparison of the different ESIMS signal intensities for [(*Isop*)$_n$.*H*]$^+$ species in experiments (A), (B), and (C) (Figure 2).** The observed signal intensities were surprisingly the strongest in the case of the gas-liquid collisions between *Isop*(g) and acidic water (pH = 1) (**C**), followed by pure isoprene (**B**), and the organic phase from emulsions of liquid isoprene and acidic water (pH = 1) (**A**). We consider that during the vigorous shaking of the isoprene-water emulsions, a fraction of the aqueous content was dissolved into the organic phase, i.e. isoprene. Subsequently, as the organic phase was electrosprayed after the phase separation, the aqueous components decreased the intensity of the isoprene oligomers due to the competing effect of the ions being attracted into the mass spectrometer. Since our ESIMS has a minimum detection limit of *m/z* = 50, we were unable to observe those smaller ions. The higher signal intensities (maximum signal intensity 68,000 a.u.) of the gas-liquid collision (**C**) could be attributed to the considerably higher flow rate of isoprene, 0.48 g/min in 600 mL/min of air, into the atmospheric chamber of the ESIMS, compared to 10 μL/min in experiments (**A**) and (**B**).

**Section S_b:**

**Determining if the mass spectrometric peaks (Figures 2, S3) correspond to covalently bonded oligomers or physisorbed clusters of isoprene.** The temperature of the glass capillary at the inlet of the mass spectrometer is an important variable in ESIMS. It provides thermal energy to the incoming droplets, facilitating the evaporation of the solvent or neutral molecules – at elevated temperatures, the non-covalently bonded clusters become unstable and the molecules separate, whereas covalently bonded oligomers survive. Thus, based on the changes in the mass spectral intensities as a function of the glass capillary temperature, we could discern if the peaks comprised of covalently bonded or non-covalently bonded species. For instance, while injecting pure acetone, ethanol, and isoprene (Figure S1 – B1, B2, and B3, respectively), we observed peaks in the spectra corresponding to $(M)_n$ ($m/z = n.M + H^+$), i.e., $(Ace)_2$, $(Et)_2$, $(Et)_3$, and $(Isop)_{2,3,4...}$ (Figure S1 – top right corner insets). However, as we increased the temperature of the glass capillary, we noticed that the intensities of the heavier peaks of acetone and ethanol dramatically decreased with increasing temperatures (Panels B1 and B2 in Figure S1). In stark contrast, the mass spectral intensities for isoprene did not vary much as the temperature increased (Panel B3 in Figure S1). Thus, our simple experiment helped us conclude that while acetone and ethanol form non-covalent clusters[1] with an excess proton, isoprene forms covalently bonded oligomers. Our conclusion was further corroborated by the clean spectra of acetone and ethanol, evidencing that negligible fragmentation and reactions were present.

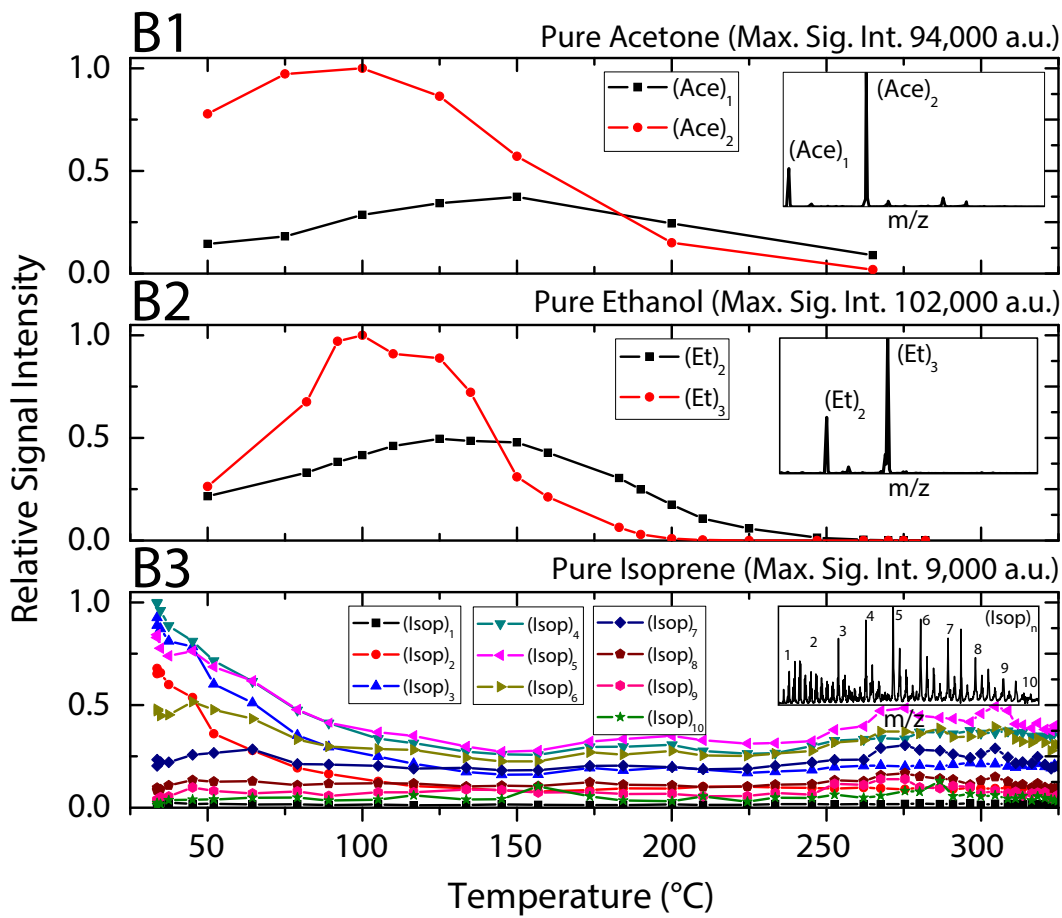

**Figure S1 – Influence of capillary temperature on the detection of [(*M*)$_n$+*H*]$^+$ peaks at the ESIMS. (B1)** Comparison between signal intensities of peaks from (*Ace*)$_1$ (m/z 59 = MW$_{ACETONE}$ + MW$_H^+$) and (*Ace*)$_2$ (m/z 117 = 2.MW$_{ACETONE}$ + MW$_H^+$). **(B2)** Comparison of peaks from (*Et*)$_2$ (m/z 93 = 2.MW$_{ETHANOL}$ + MW$_H^+$) and (Et)$_2$ (m/z 138 = 3.MW$_{ETHANOL}$ + MW$_H^+$). **(B3)** Comparison of peaks from (*Isop*)$_1$ (m/z 69 = 1.MW$_{ISOPRENE}$ + MW$_H^+$) and (*Isop*)$_{10}$ (m/z 681 = 10.MW$_{ISOPRENE}$ + MW$_H^+$). ESIMS set to positive mode, 6 kV, 10 µL/min for acetone and ethanol, and 5 µL/min for isoprene. The insets in the right corners are characteristic ESIMS spectra for each case.

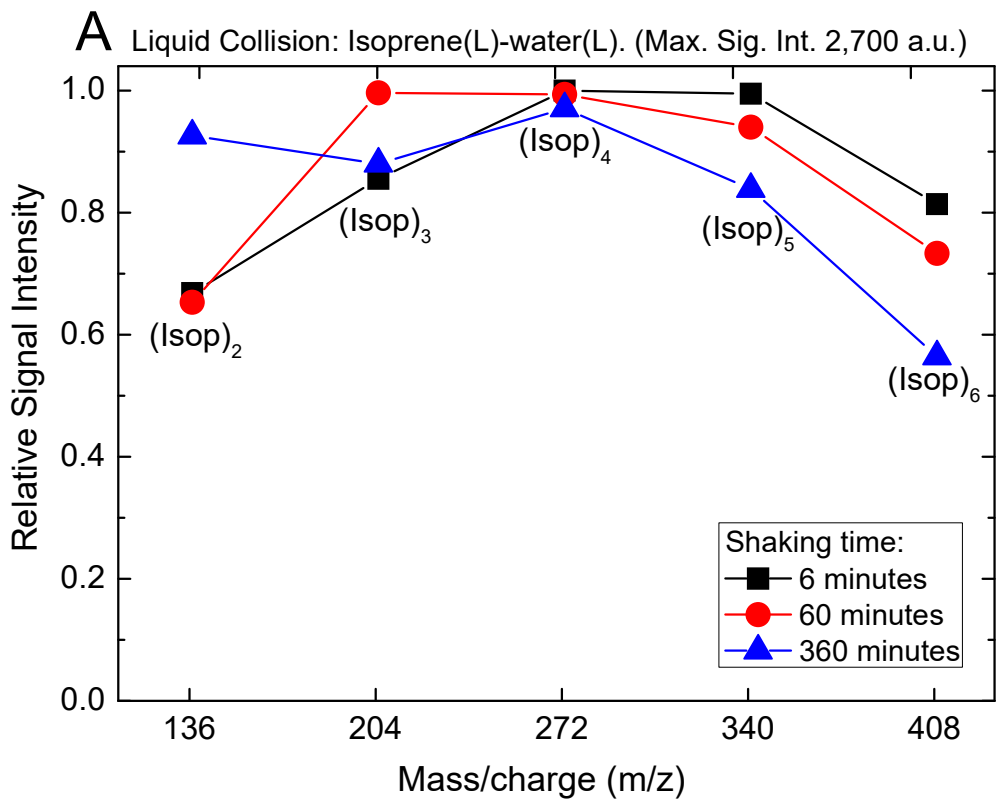

**A** Liquid Collision: Isoprene(L)-water(L). (Max. Sig. Int. 2,700 a.u.)

**Figure S2 - Influence of shaking duration on the liquid-liquid collisions, experiments (A).** Comparison of ESIMS peaks from (*Isop*)$_2$ (m/z 136 = 1.MW$_{ISOPRENE}$ + MW$_H^+$) and (Isop)$_6$ (m/z 409 = 6.MW$_{ISOPRENE}$ + MW$_H^+$); ESIMS set to positive mode, 6 kV, 150 °C, 10 μL/min, pH of aqueous phase 1.52. The shaking time did not considerably influence the signal intensity of the liquid-liquid collisions.

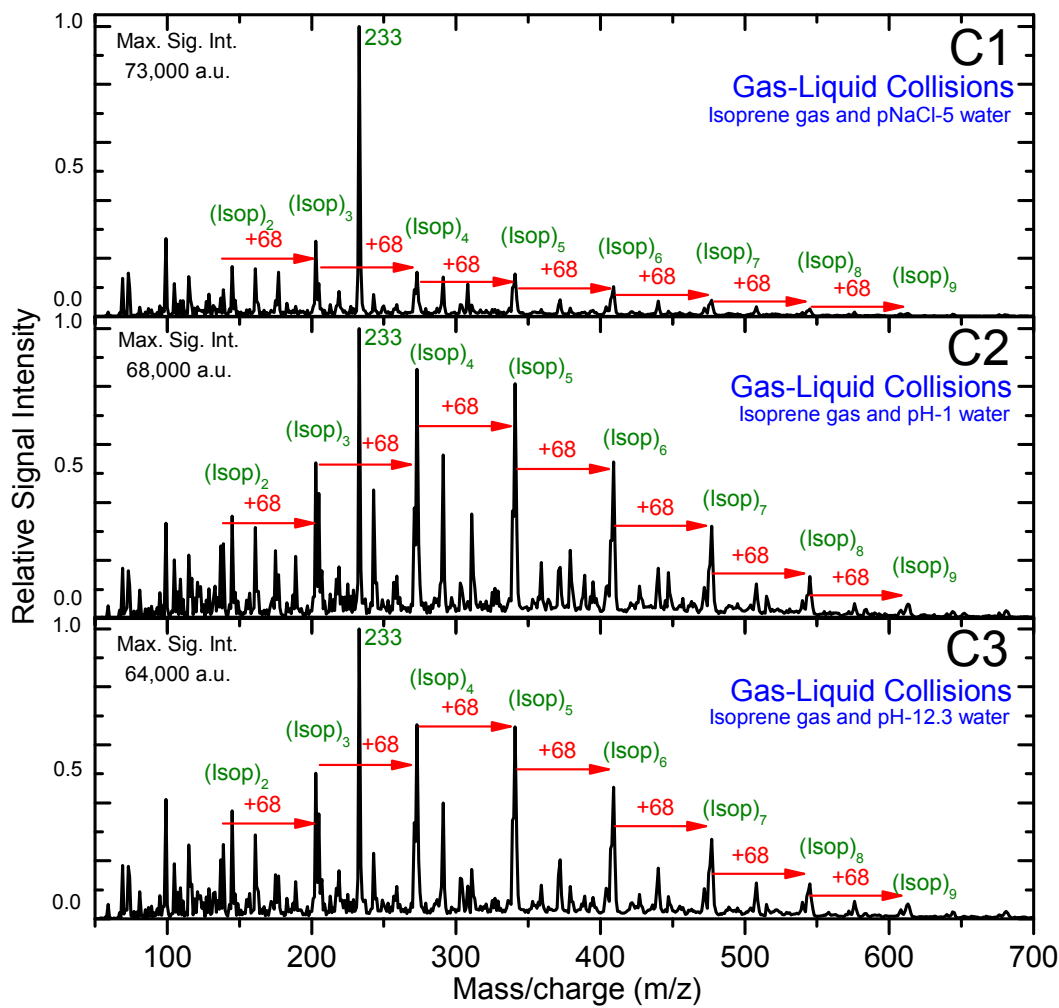

**Figure S3 – ESIMS spectra for gas-liquid collisions, experiments (C).** (**C1**) Water at p[NaCl] = 5 and flow rate of 1 μL/min; (**C2**) Water at pH = 1 and flow rate of 1 μL/min; (**C3**) Water at pH = 12.3 and flow rate of 1 μL/min. The main peaks correspond to the oligomers of isoprene plus one proton [(*Isop*)$_n$+*H*]$^+$, and are separated by the mass of isoprene (68 *m/z*). The four smaller peaks between the main ones correspond to the four possible fragmentations of the carbon bonds within the isoprene molecule (C$_5$H$_8$). In all three cases the stream flow of air (600 mL/min) and gaseous isoprene (which evaporated from the air bubbler at ~0.48 g/min) was directed towards the electrosprayed water jet; the temperature of the capillary in the ESIMS was 150 °C, the electrical potential applied at the electrospray needle was 6 kV, the capillary inlet was grounded, and the separation between the needle and the capillary inlet was ~1 cm.

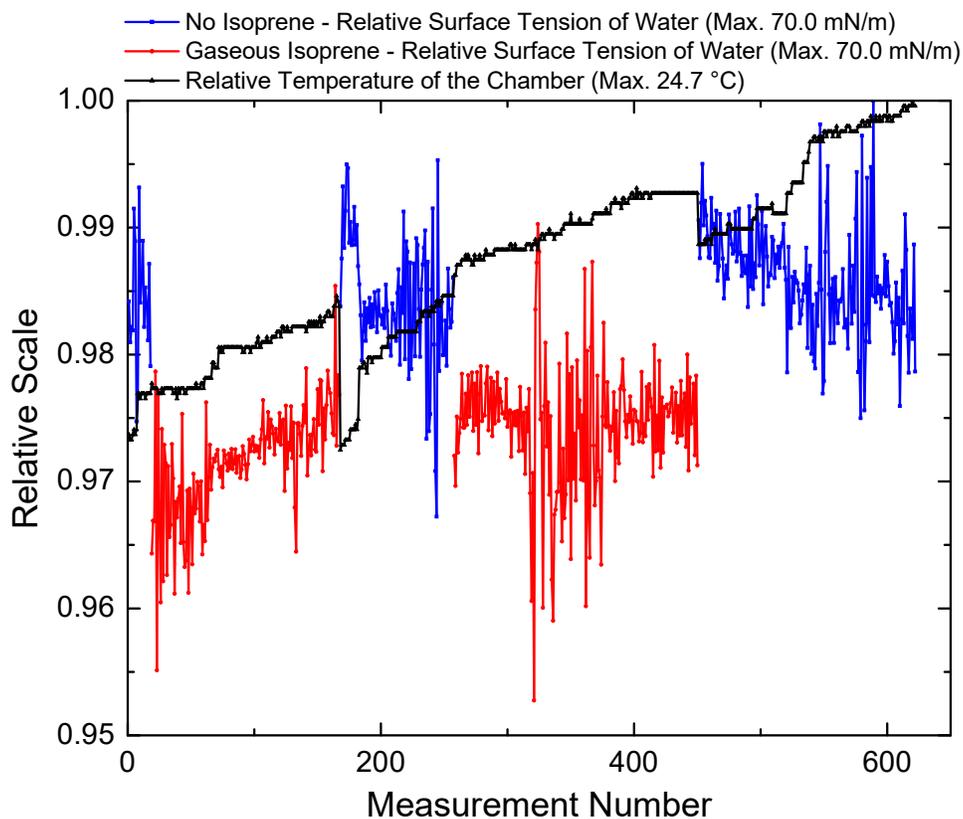

**Figure S4 – Influence of gaseous isoprene on the surface tension of a droplet of water.** The surface tension was measured by the pendant drop method in a chamber in the presence of air (blue line) and gaseous isoprene in air (red line). The surface tension of the water droplet changed considerably in the presence of gaseous isoprene, indicating adsorption and possibly dissolution of the isoprene into the bulk water.

**Section S$_c$:**

  **What is the role of the net positive charge during droplet formation at the electrospray?** Let us consider that the charge separation (and oxidation reactions) at the metallic needle ejecting an analyte under electric fields leads to a decrease in the pH of the just-formed droplets by at least 0.1 (or a 26% increase in the concentration of protons)[2, 3]. The initial droplet size will depend on the electric potential, polarity, needle diameter, ionic strength, surface tension, and viscosity (Figure S5 contains a visual representation of this thought experiment). While evaporating, the charge density of the droplets increases and the Rayleigh limit is eventually reached, i.e. the repulsion of the electrostatics will overpower the cohesion of the surface tension. The excess charge is predicted to be $Q = (k\pi^2\varepsilon_0\gamma R^3)^{1/2}$, where $d$ is the diameter of the droplet at the time of fission, $\gamma$ is the surface tension of the liquid, $\varepsilon_0$ is the permittivity of a vacuum, and $k$ is a constant[4, 5].

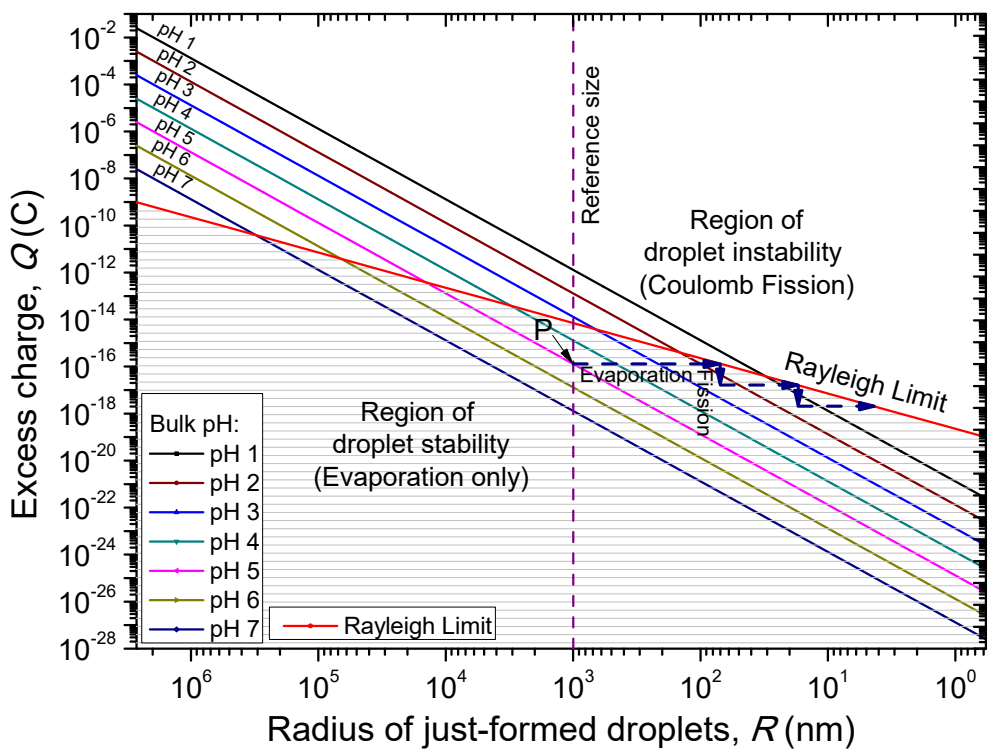

**Figure S5 – Simulation of the influence of bulk pH and initial radius on the net/excess charge of electrosprayed droplets.** For a given ESI setup, the just-formed droplets will have different diameters depending on experimental conditions such as the electric potential, polarity, needle diameter, ionic strength, and surface tension. We expect that the sooner the droplets reach the Rayleigh limit, the faster they will undergo Coulomb fissions and release highly reactive clusters containing excess hydronium ions. Let us consider water with pH = 5 (pink line): a droplet of 1000 nm is formed with an excess charge of ~26%, i.e. $10^{-16}$ C (Point $P$).

Since this droplet is below the Rayleigh limit, it will not undergo Coulomb fission immediately. However, this pH 5 droplet will evaporate (following the horizontal arrow) and then, after reaching the Rayleigh limit, undergo Coulomb fission. Note that the down arrow representing fission is exaggerated; it shows a ~10-fold decrease in the droplet charge during each fission, while a lower discharge is expected in reality[3]. Conversely, droplets with pH = 1 (black line) would immediately eject hydroniums to the gas phase under same setup conditions. A parallel can be drawn between this process and the proposed Mechanism $M_3$.

**Section S_d:**

**Computational Section:** We carried out density functional theory (DFT) calculations at the M06 level to provide an accurate description of the ground-state thermochemistry and thermochemical kinetics of the isoprene (*Isop*) and water clusters[6-9]. The calculated transition state structures and energies of a series of organic reactions with M06 are in good agreement with the experimental data[8]. Researchers have evaluated the binding energies of water clusters, $(H_2O)_n$ (range $n$ = 2-8, 20), as well as the hydration and neutralization energies of hydroxide and hydronium ions using DFT functionals (M06, M06-2X, M06-L, B3LYP, X3LYP), and compared these energies against high-level theory (CCSD(T)/aug-cc-p VDZ level)[6]. They found the results from M06 to be in excellent agreement with the high-level theory, both with and without the basis set superposition error correction.

*Ab initio* predictions of the proton transfer thermodynamics between $H_3O^+(g)$ and *Isop*(g) were $\Delta G^0$ = -30.7 kcal mol$^{-1}$, in accordance with the experimental gas-phase basicities (GB) of $H_2O$ ($GB_{H2O}$ = -157.7 kcal mol$^{-1}$) and *Isop* ($GB_{ISO}$ = -190.6 kcal mol$^{-1}$), $\Delta GB$ = -32.9 kcal mol$^{-1}$ (Figure S6)[10]. Furthermore, the trans- or cis-Isop(g) spontaneously adds to (Isop.$H^+$ + $H_2O$), leading to cyclic ($\Delta G^0$ = - 40 kcal mol$^{-1}$) or acyclic monoterpenes ($\Delta G^0$ = - 9 kcal mol$^{-1}$) (Figures S7 and S8), as noted by other researchers[11, 12]. We simulated small clusters of water with an excess proton produced during the ESIMS processes as $(H_2O)_3.H^+$. Since the proton was insufficiently hydrated in this cluster, it exhibited extreme acidity. The incipient isoprene molecule (*ISOP*) fell into a shallow potential well forming an adduct (Figure 5). The free energy barrier for proton transfer from $(H_2O)_3.H^+$ to *ISOP*(g) was $\Delta G^{\ddagger}$ = 6.9 kcal mol$^{-1}$; the barrier to subsequent oligomerization with another free ISOP(g) was $\Delta G^{\ddagger}$ = 2.1 kcal mol$^{-1}$. As shown in Figure 6 of the manuscript, the kinetic barriers for protonation and oligomerization of isoprene on a larger $(H_2O)_{36}.H^+$ cluster were $\Delta G^{\ddagger}$ = 25.5 kcal mol$^{-1}$ and $\Delta G^{\ddagger}$ = 40.2 kcal mol$^{-1}$, respectively, which are insurmountable under ambient conditions within a 1 ms time frame. Thus, our models are representative of the high-energy ESIMS processes at extended air-water interfaces under equilibrium conditions.

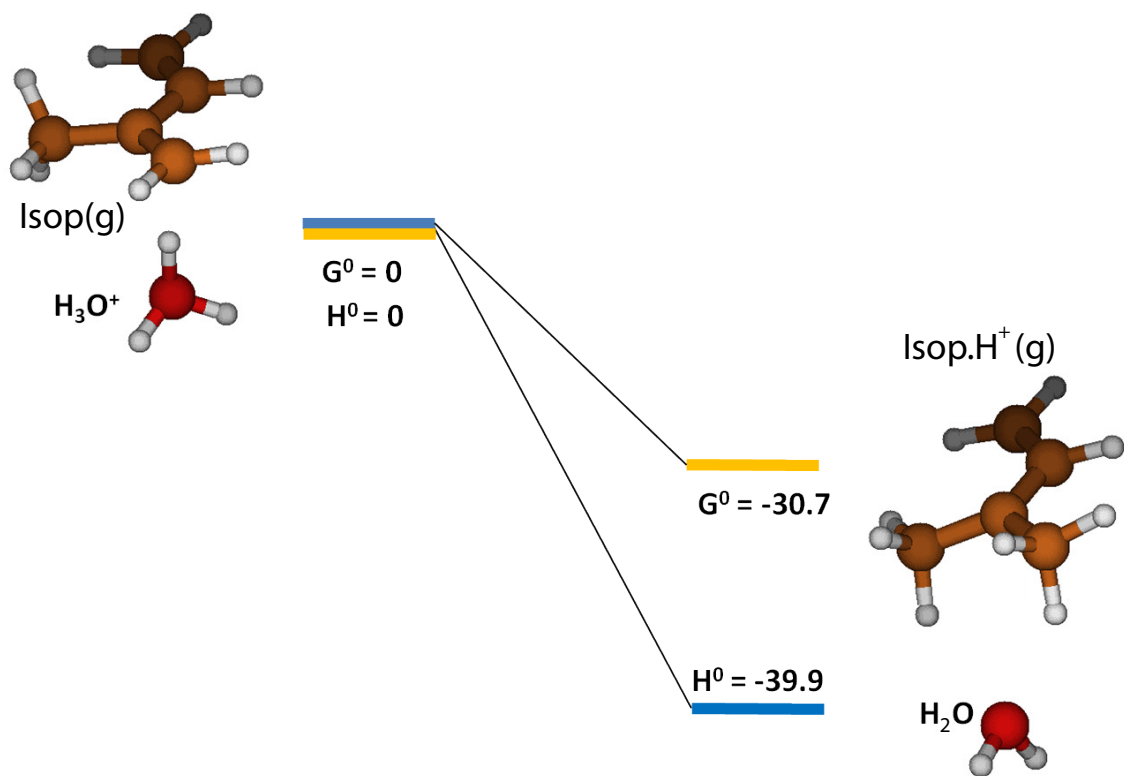

Isop(g)

$H_3O^+$

$G^0 = 0$
$H^0 = 0$

Isop.H$^+$(g)

$G^0 = -30.7$

$H^0 = -39.9$

$H_2O$

**Figure S6 -** *Ab initio* **predictions of the proton transfer reaction between a gas-phase hydronium ion, $H_3O^+$(g), and a gas-phase isoprene molecule, *Isop*(g).** Theory predicted the reaction to be spontaneous with a free energy change of $\Delta G^0 = -30$ kcal mol$^{-1}$, in accordance with the experimental gas-phase basicities (GB) of $H_2O$ (GB$_{H2O}$ = 157.7 kcal mol$^{-1}$) and *Isop* (GB$_{ISO}$ = 190.6 kcal mol$^{-1}$); $\Delta$GB = 32.9 kcal mol$^{-1}$.

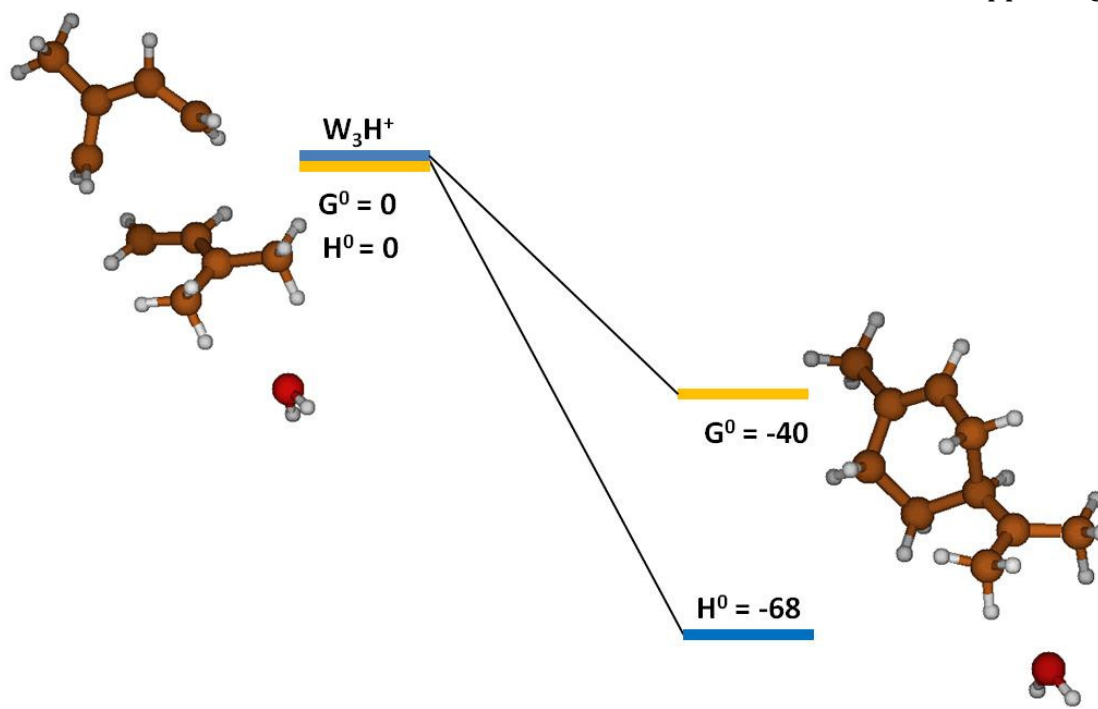

**Figure S7 - Spontaneous gas-phase oligomerization of gas-phase cis-isoprene (*Isop*(g)) with a protonated trans-isoprene molecule, leading to a cyclic product.** The units for $G^0$ and $H^0$ are kcal-mol$^{-1}$.

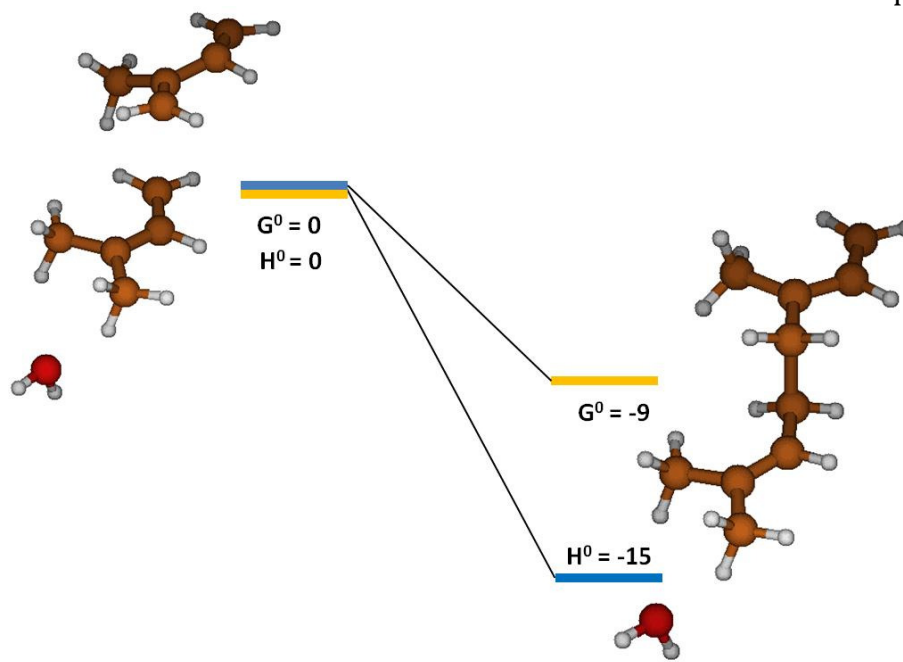

**Figure S8 – Spontaneous gas-phase oligomerization of a trans-isoprene (*Isop*(g)) molecule with a protonated trans-isoprene, leading to a linear product.** The units for $G^0$ and $H^0$ are kcal-mol$^{-1}$.

**Section S$_e$:**

**Description of the NMR:** NMR signals showed at the spectra obtained in Liquid-liquid collisions experiments (A) and pure, as-purchased isoprene (B):

**$^1$H NMR** (700 MHz, Chloroform-*d*) **δ (ppm):** 6.47 (dd, *J* = 17.5, 10.8 Hz, 1H, **H2**), 5.20 (d, *J* = 17.5 Hz, 1H, **H1b**), 5.09 (d, *J* = 10.8 Hz, 1H, **H1a**), 5.02 (d, *J* = 13.2 Hz, 2H, **H5b and H5b**), 1.87 (s, 3H, C**H$_3$**).

**$^{13}$C NMR** (176 MHz, Chloroform-*d*) **δ (ppm):** 142.29 (**C3**), 139.61 (**C2**), 116.76 and 113.65 (**C1** and **C5**), 17.80 (**C4**).